\documentstyle[aps,preprint]{revtex}

\begin{document}
\title{\bf A Unified Model for Two Localisation Problems: Electron
States in Spin-Degenerate Landau Levels, and in a Random Magnetic Field} 
\author{D.K.K. Lee\cite{usa} and J.T. Chalker}
\address{Theoretical Physics, University of Oxford, Keble Road, Oxford, OX1
3NP, United Kingdom}

\maketitle

\begin{abstract}
A single model is presented which represents both of the two
apparently unrelated localisation problems of the title. The phase
diagram of this model is examined using scaling ideas and numerical
simulations. It is argued that the localisation length in a
spin-degenerate Landau level diverges at two distinct energies, with
the same critical behaviour as in a spin-split Landau level, and that
all states of a charged particle moving in two dimensions, in a random
magnetic field with zero average, are localised.
\end{abstract}
\bigskip
\pacs{PACS numbers: 72.10.Bg,71.55.Jv,72.20.My}
\tighten

Two apparently disparate localisation problems of great current
interest arise in the context of the quantum Hall effect. The first
concerns the delocalisation transition in Landau levels that are
spin-degenerate, while the second involves localisation in the
presence of a random magnetic field. In this paper we show that both
physical situations can be represented by a single model, and that
conclusions about one carry implications for the other. We present the
results of extensive numerical simulations, which elucidate the
behaviour of our unifying model.

The mobility edge in {\it spin-split} Landau levels represents the
best characterised example to date of critical behaviour at a
metal-insulator transition. Interpretation of the experiments starts
from a picture for the dependence on energy, $E$, of eigenstate
properties, which is derived from scaling ideas
\cite{levine83} and supported by numerical simulations
\cite{aoki85,chalker88,huck90,bhatt}. In this picture, almost all
eigenstates within a disorder-broadened Landau level are Anderson
localised, but the localisation length, $\xi(E)$, diverges at a
critical energy, $E_c$, which, in the simplest case, lies at the band
centre: $\xi(E) \sim |E - E_c|^{-\nu}$. Scaling has been observed as
this zero-temperature critical point is approached, as a function of
temperature \cite{wei88}, of sample size \cite{kochnew} and of
frequency \cite{engel}. Early measurements of the ratio $\kappa=z/\nu$,
where $z$ and $\nu$ are the dynamic scaling exponent and the
localisation length exponent, respectively, obtained the value
\cite{wei88}, $\kappa \simeq 0.42$, while subsequent experiments on
mesoscopic Hall bars have allowed $z$ and $\nu$ to be determined
independently, yielding the value \cite{kochnew} $\nu=2.3\pm0.1$, in
excellent accord with the results of numerical simulations using a
variety of models and techniques \cite{chalker88,huck90,bhatt}.

By contrast, in sufficiently disordered samples at low magnetic field,
it is possible for the disorder-broadening of Landau levels to exceed
the Zeeman energy, so that each orbital level becomes {\it
spin-degenerate}. Under these conditions different critical behaviour
is reported \cite{wei88,kochnew,engel}. Assuming, as for spin-split
Landau levels, that the localisation length diverges only at one
energy, a value of the ratio $\kappa$ smaller by a factor of $2$,
$\kappa\simeq 0.21$, is obtained from scaling both with temperature
\cite{wei88} and frequency \cite{engel}. Measurements on mesoscopic
samples with spin-degenerate Landau levels are in qualitative accord,
giving a larger value \cite{kochnew} for $\nu$, $\nu \simeq 6.5$.

We present below results of numerical simulations that are consistent
with these observations. Supposing, as in the experimental analysis,
that the localisation length diverges only at the centre of a
spin-degenerate Landau level, we obtain, for localisation lengths in
our model similar in magnitude to those probed experimentally, a much
faster divergence than in spin-split Landau levels, with $\nu \simeq
5.8$ as the best fit under these assumptions. We argue, however, that
such an analysis is, in fact, incorrect. We believe, instead, that
there are two energies at which the localisation length diverges in a
spin-degenerate Landau level, displaced symmetrically by small amounts
to either side of the Landau level centre, and that the localisation
length between these mobility edges is large but not infinite. Indeed,
re-analysing out numerical data under this new assumption, we obtain a
substantially better fit to a power-law divergence of the localisation
length, with the same exponent value, $\nu \simeq 2.3$, as found in
spin-split Landau levels.

Interest in localisation in two dimensions, in the presence of a random
magnetic field with an average value of
zero, arises both from the Chern-Simons theory of electrons in a
half-filled Landau level \cite{kalmeyer,read} and from
gauge theories of doped Mott insulators \cite{nagaosa}. Recent simulations of
localisation in a random magnetic field have resulted in conflicting
interpretations, some authors \cite{kalmeyer,arovas} arguing that
there exists a range of energies for which states are extended, while
others \cite{sugiyama} suggest that all states are localised.

The data we present below indicate that in our model for localisation
in a random magnetic field, all states are localised. Moreover, we
argue that should extended states exist in this problem, it would have
implications for the quantum Hall effect that would require revision
of the usually accepted scaling flow diagram \cite{levine83}.

Our calculations use an extension of the `network model', introduced
previously by one of us \cite{chalker88}. As a representation of a
spin-split Landau level, the original network model is based on a
semiclassical picture of electron motion in a smooth, two-dimensional
random potential, under a strong, perpendicular magnetic field. Taking
the magnetic length to be much shorter than the correlation length of
the potential, electron motion can be separated into two components: a
rapid cyclotron orbit and a slow drift of the guiding centre along
equipotential lines. At this level, the spatial extent of eigenstates
depends only on classical percolation properties of the equipotential
lines.  Quantum effects are introduced by allowing tunnelling between
disjoint portions of the same equipotential, near saddle points of the
potential. Electrons propagate coherently through the system, so that
interference between different tunnelling paths is automatically
included.

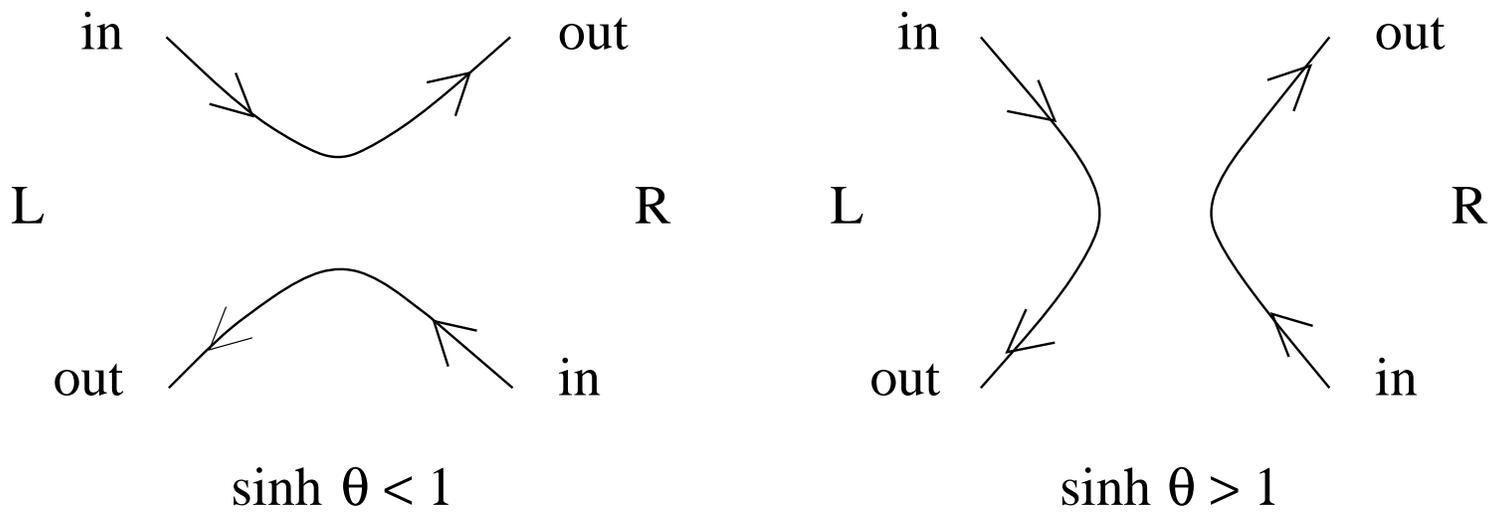
\begin{figure}[hbt]
\caption{Node parameters. Diagrams indicate tendency for transmission
and reflection for different node parameters $\theta$.}
\label{node}
\end{figure}

Formally, the model for a spin-split Landau level consists of a set of
links, representing portions of an equipotential, which meet at nodes,
representing saddle points. For simplicity, the links and nodes are
arranged on a square lattice. Each link is characterised by a
direction for flow of probability flux, which is the direction of the
corresponding guiding centre drift, and by the phase shift that an
electron acquires on traversing the link. Randomness is introduced by
choosing these link phases from a uniform distribution. Each node is
described by a scattering matrix which, after allowing for the
constraints of unitarity, contains only one important quantity, the
`node parameter', $\theta$. Current amplitudes on the four links
meeting at a given node are related by (referring to Fig. \ref{node})
\begin{equation}
\left( \begin{array}{c} \psi_{\rm in,R} \\ \psi_{\rm out,R} \end{array} \right) =
\left( \begin{array}{cc} \cosh \theta & \sinh \theta \\ \sinh \theta 
& \cosh \theta \end{array} \right)
\left( \begin{array}{c} \psi_{\rm out,L} \\ \psi_{\rm in,L} \end{array} \right)
\end{equation}
where, for brevity, we have chosen a gauge in which all amplitudes
\{$\psi$\} have the same phase. Tunnelling is turned off in two
limits: $\theta = 0$, when $\psi_{\rm out,R} = \psi_{\rm in,L}$ and
$\psi_{\rm in,R}=\psi_{\rm out,L}$; and $\theta \rightarrow \infty$,
when $\psi_{\rm in,R}=\psi_{\rm out,R}$ and $\psi_{\rm
out,L}=-\psi_{\rm in,L}$. More generally, it is useful to note the
following duality transformation: if the amplitudes $\psi_{\rm out,R}$
and $\psi_{\rm out,L}$ are permuted in (1), $\theta$ should be
replaced by $\theta^{\prime}$, where $\sinh
\theta^{\prime} = 1/ \sinh \theta$.  Thus tunnelling is maximal at
$\sinh \theta = \sinh \theta^{\prime} = 1$.

In the generalisation of this network model which we study below, two
quantum-mechanical fluxes are carried by each link.  Scattering
between these two channels is included by replacing the link phases of
the original model with $U(2)$ matrices that transform the two
incident current amplitudes on a link into outgoing ones. We choose
these matrices randomly and independently on each link, with the Haar
measure. At the nodes, we suppose (without loss of generality
\cite{hopefully}) that tunnelling conserves the channel index, with
parameters $\theta_1$, $\theta_2$ for each channel respectively. For
simplicity, we do not consider randomness in $\theta_1$ or $\theta_2$. 

As a representation of a spin-degenerate Landau level, the two
channels correspond to the two possible spin orientations and the
non-zero U(2) mixing arises from spin-orbit scattering. In the absence
of Zeeman splitting, $\theta_1=\theta_2\equiv \theta$, and on sweeping
the Fermi energy through the Landau level, the system follows a line
in parameter space from $\theta=0$ to $\theta=\infty$. With Zeeman
splitting, this line is displaced: schematically, one can take
$\theta_1 = (1+g)\theta$ and $\theta_2 = (1-g)\theta$, where $g$
represents the electron $g$-factor, with $|g|<1$.

The same model also represents electron motion in a random magnetic
field that has a correlation length large compared to the typical
cyclotron radius. In this semiclassical limit it is again useful to
consider electron guiding centres, which under these conditions drift
along contours of the magnetic field. Extended states, if any exist,
must be associated with the percolation of the zero field contour.
Drift along the zero field contour has been discussed in detail in
Refs \cite{muller,chklov93}. A mapping to our network model is
most easily established by considering a special case, in which the
field switches abruptly between two values, $\pm B_0$, as the contour
is crossed, and by restricting attention to Fermi energies lying
between the energies of the lowest two Landau levels in a uniform
field of strength $B_0$.  Generalisations will be treated elsewhere
\cite{hopefully}. In these circumstances \cite{chklov93}, two modes
propagate along the contour in the same direction, arising ultimately
from symmetric and antisymmetric linear combinations of the lowest
Landau levels on either side. In our model, a portion of this contour
corresponds to a link, which hence must support two channels. Where
two different portions of contour approach within a magnetic length of
each other, tunnelling can occur between them, as described by the
nodes of the model. Moreover, meandering of the contour will result in
scattering between the modes, represented by the $U(2)$ matrices. We
identify the line in the parameter space $(\theta_1,\theta_2)$ which
has the symmetry of the random field problem from the condition that
the average Hall conductance vanish: in order that the average
circulation around plaquettes of the network model be zero, we require
$\sinh
\theta_1 = 1/ \sinh \theta_2$ \cite{footnote1}.

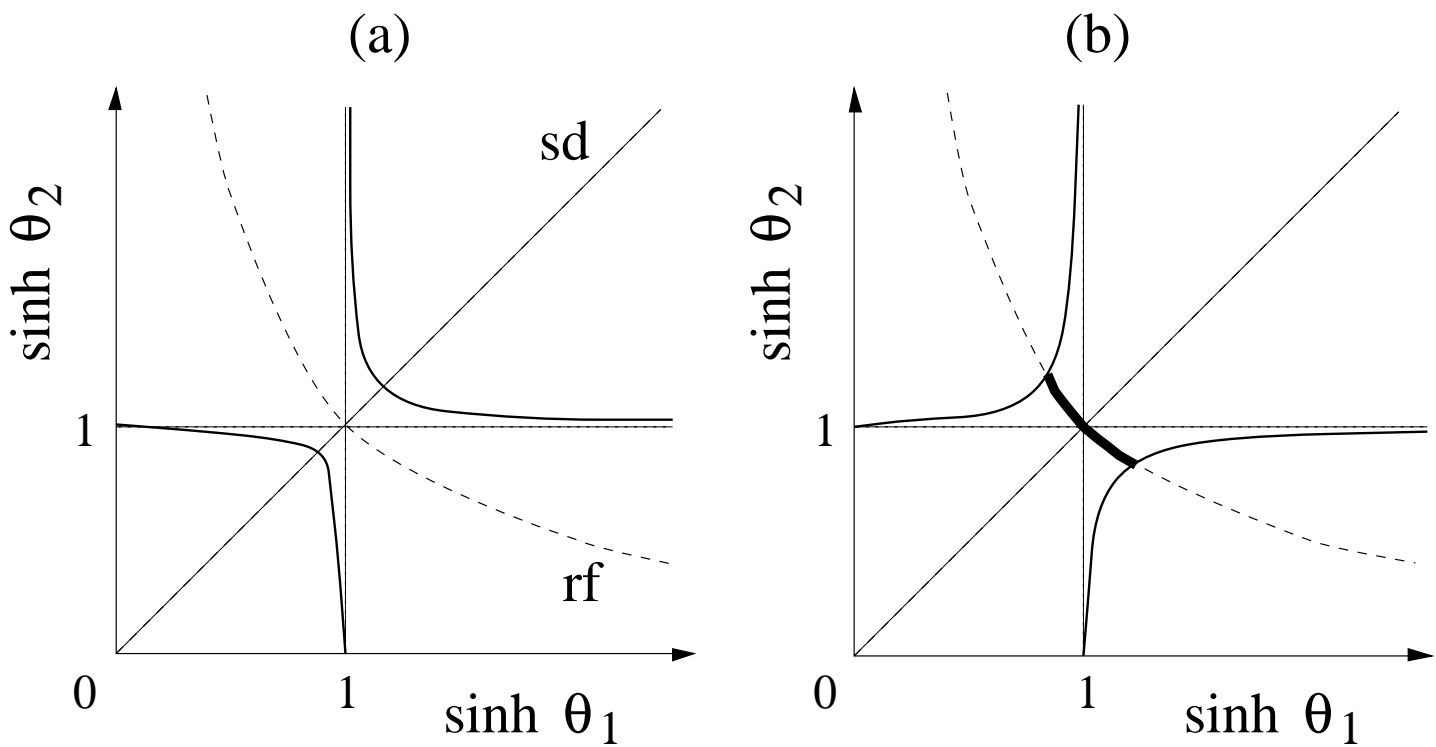
\begin{figure}
\caption{Schematic phase diagrams: (a) supported by our results (b)
alternative with unconventional critical line (bold line). (sd: spin-degenerate
network, rf: random field network)}
\label{phase}
\end{figure}

We are now in a position to discuss the phase diagram for both
problems in terms of our unifying model. In the absence of scattering
between channels, the model would consist of two uncoupled networks,
each as studied in Ref \cite{chalker88}. States are localised, except
(in the respective networks) on the lines $\sinh \theta_1 = 1$ and
$\sinh \theta_2 = 1$. In this paper, we are concerned with the result
of coupling the two networks. First, consider, as in Ref \cite{khm},
traversing the spin-degenerate Landau level on the line: $\theta_1 =
\theta_2 \equiv \theta$. Along this line, the Hall conductance,
$\sigma_{\rm xy}$, measured at short distances in units of $e^2/h$,
must vary smoothly between $\sigma_{\rm xy}=0$ at $\theta=0$ and
$\sigma_{\rm xy}=2$ as $\theta\to\infty$. In particular, $\sigma_{\rm
xy}=1$ at $\sinh \theta_1 = \sinh
\theta_2 = 1$, which is the centre of the spin-degenerate Landau level, and
the Khmelnitskii flow diagram \cite{levine83} suggests that scaling
takes the system to a localisation fixed point. Additionally, one
expects \cite{khm} to find two isolated points either side of the
Landau level centre, with $\sigma_{\rm xy}=1/2$ and $3/2$
respectively, from which the system scales towards a delocalisation
fixed point. Since any trajectory from $\theta_1 =
\theta_2 = 0$ to $\theta_1 = \theta_2 = \infty$ must, on this
analysis, share these features, one is led to the phase diagram of
Fig. \ref{phase}(a). In this phase diagram, there are two distinct mobility
edges in the spin-degenerate Landau level and all states on the random
field line are localised. An alternative scenario requires that two
delocalisation fixed points in the scaling flow diagram coalesce if
the Zeeman energy is small enough, and yields the phase diagram of
Fig. \ref{phase}(b). In this event, states are delocalised only at the {\it
centre} of the spin-degenerate Landau level (with, potentially,
critical properties in a new universality class) and there exists an
entire region of extended states on the random field line. Clearly,
one can also imagine more exotic possibilities, which we do not
discuss.

To identify which phase diagram actually pertains, we have carried out
numerical simulations, employing standard techniques (see
\cite{chalker88} and references therein). We measure the localisation
length $\xi_M$ in networks which are up to $M=128$ links wide and
$1.2\times10^6$ links long, using periodic boundary conditions to
avoid extended edge states. Localisation lengths in the thermodynamic
limit, $\xi_\infty$, are extracted via a conventional one-parameter
scaling analysis.

We discuss first our results on the random field line, $\sinh \theta_1
= 1/\sinh \theta_2$. The localisation length remains finite everywhere
on this line (Fig. \ref{rndf}), including, notably, the point, $\sinh
\theta_1 =\sinh \theta_2 = 1$, that coincides with the spin-degenerate
Landau level centre. Such behaviour is expected from the phase diagram
of Fig. \ref{phase}(a), but is incompatible with that of Fig. \ref{phase}(b).

Consider next a Landau level without Zeeman splitting: $\theta_1 =
\theta_2 \equiv \theta$. On this line, over a narrow range on either side
of the level centre, the bulk localisation length is so large that we
are unable to determine it reliably. The divergence in the
localisation length, as the level centre ($\sinh
\theta = 1$) is approached from the low energy tail ($\theta \ll 1$)
is examined in Fig. \ref{so}, comparing the two alternative hypotheses
represented by Fig. \ref{phase}. According to
the former, one must simultaneously determine the position,
$\theta_{\rm c}$, of the lower mobility edge and the value, $\nu$, of the
critical exponent. Since large uncertainties are associated with such
a two-parameter fit, we simply demonstrate that there exists a choice
for $\theta_{\rm c}$ for which the data are consistent with the exponent
value, $\nu = 2.3$, obtained in simulations
\cite{chalker88,huck90,bhatt} of a spin-split Landau level. Supposing,
alternatively (as in the experimental analysis \cite{wei88,kochnew},
but in contradiction to Fig. \ref{rndf}), that there is a mobility edge at the
level centre, our data fit less well to a power law and require a much
larger exponent, $\nu \simeq 5.8$, in at least qualitative accord with
experiment \cite{wei88,kochnew}. We suggest that it would be of
considerable interest to analyse experiments on spin-degenerate Landau
levels with the assumption that there are two distinct mobility edges,
as in Fig. \ref{phase}(a).

Finally, we attempt a precise determination of the critical exponent,
$\nu$, in our model. To do so, we study the line $\sinh \theta_2 = 2$,
which is equivalent to introducing Zeeman splitting. In addition, in
order to reduce the localisation length to measurable values in the
range $\sinh \theta_1 \approx 1$, it is necessary to decrease the
coupling between channels on each link. We achieve this by abandoning
an isotropic distribution for the $U(2)$ matrices, and restricting
their off-diagonal elements to have modulus $0.3$. The localisation
length diverges as the region $\sinh \theta_1 \approx 1$ is approached
from either side. We fit power laws to each divergence, assuming a
mobility edge at $\theta_1 = \theta_{\rm c}$, obtaining exponents $\nu_-(\theta_{\rm c})$
and $\nu_+(\theta_{\rm c})$ on each side.  Solving the equation $\nu =
\nu_-(\theta_{\rm c}) = \nu_+(\theta_{\rm c})$, we obtain $\nu=2.45$  and
$\sinh \theta_{\rm c} = 1.00$. This exponent value is remarkably close to the
most precise determination for spin-split Landau levels, $\nu=2.34 \pm
0.04$ \cite{huck90}.

In summary, our numerical study has shown that, in two dimensions, all
states are localised in a random magnetic field and that the
spin-degenerate Landau levels have a pair of delocalisation
transitions in the same universality class as the spin-split system.
We have further demonstrated that these conclusions are mutually
consistent when cast in a wider parameter space containing both
quantum Hall systems of arbritrary Zeeman splitting and the random
field problem, as summarised in the phase diagram of
Fig.\ref{phase}(a).

We thank D.  Arovas, S.M.  Girvin, D.E. Khmelnitskii, D. Ko, D.-H.
Lee, N. Read and B. Shklovskii for discussions, and the Aspen Centre
for Physics, for hospitality.  This work was supported by the Science
and Engineering Research Council (GR/GO 2727) and by the European
Community (SCC CT90 0020).

\begin{figure}
\caption{Random field network. Bulk localisation length $\xi_\infty$ as a
function of $\sinh\theta$. Insets: one-parameter scaling function
$\xi_M/M = f(\xi_\infty/M)$ and schematic diagram of network.}
\label{rndf}
\end{figure}

\begin{figure}
\caption{Spin-degenerate Landau level. A better power-law fit to the
form $\xi_\infty \sim |\sinh\theta-\sinh\theta_{\rm c}|^{-\nu}$ is
obtained for $\sinh\theta_{\rm c}=1.35$. Inset: schematic diagram of
network.}
\label{so}
\end{figure}


\begin{references}
\bibitem{footnote1}At the point $\sinh \theta_1 = \sinh \theta_2 = 1$,
the line representing the random field problem crosses that
representing a spin degenerate Landau level at the mid-point in energy
of the Landau level. Here, we have the apparently paradoxical
requirement that the average Hall conductivity is zero for the former
system, but $e^2/h$ for the latter. The paradox is resolved by noting
that different boundary conditions apply in the two cases. Edge states
carry no Hall current in the first case, but contribute unit Hall
conductance in the second case. See \cite{hopefully}.
\end{references}
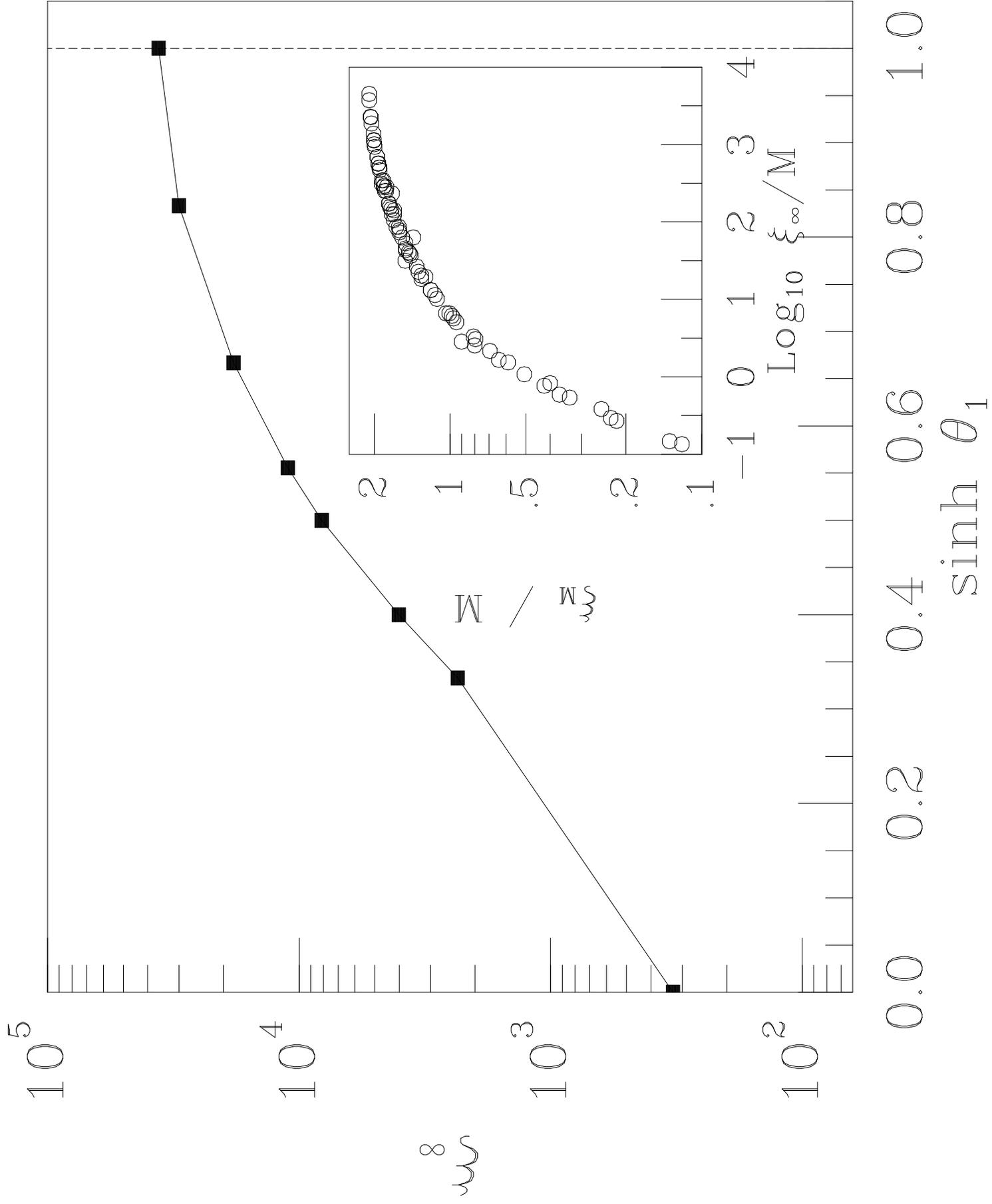
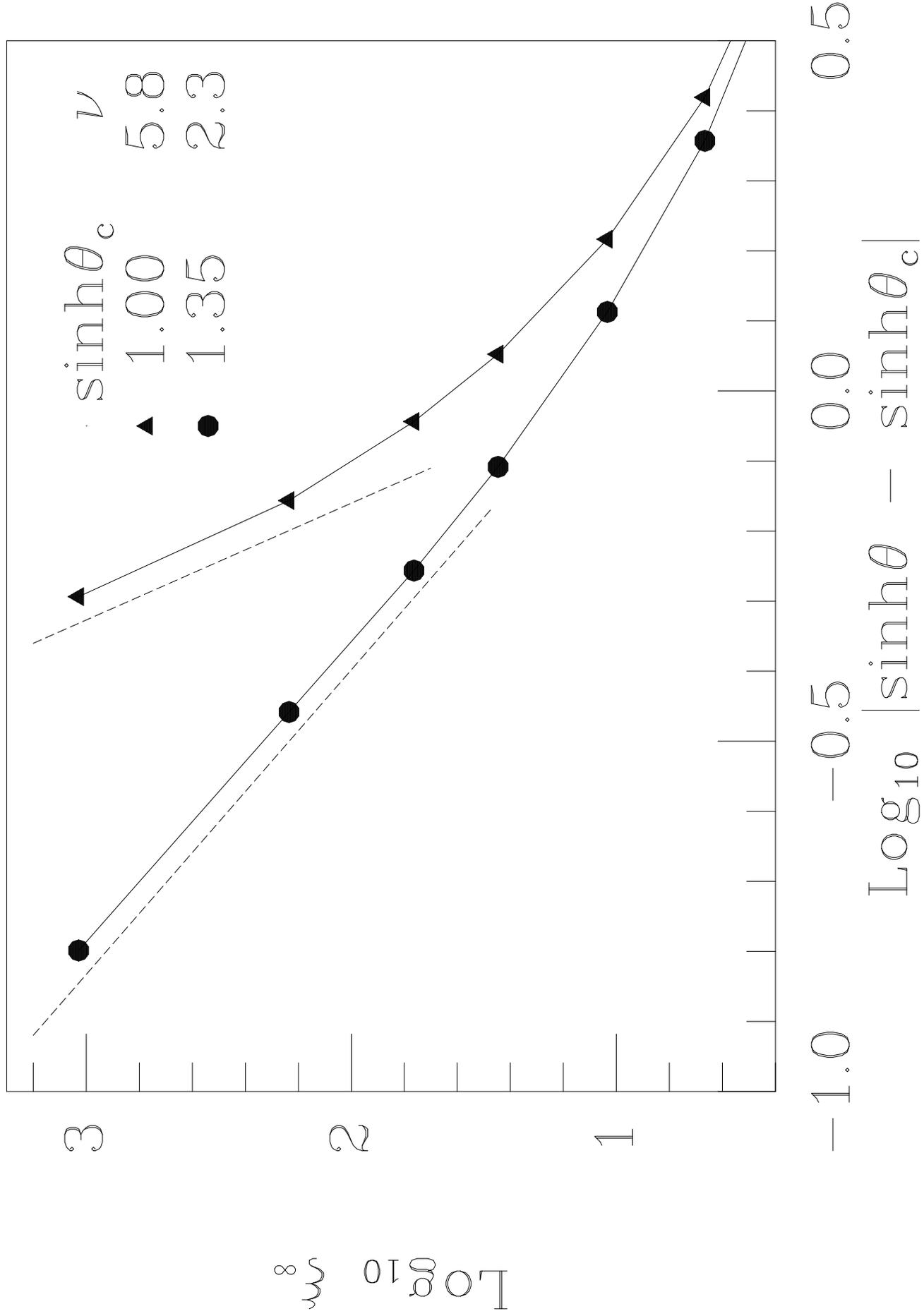
\end{document}